\documentstyle[12pt,psfig,epsf]{article}
\newcommand{\be}{\begin{equation}}
\newcommand{\ee}{\end{equation}}

\newcommand{\beqa}{\begin{eqnarray}}
\newcommand{\eeqa}{\end{eqnarray}}

\newcommand{\eqref}[1]{(\ref{#1})}

\def\boxit#1{\vbox{\hrule\hbox{\vrule\kern8pt
\vbox{\hbox{\kern8pt}\hbox{\vbox{#1}}\hbox{\kern8pt}}
\kern8pt\vrule}\hrule}}
\def\mathboxit#1{\vbox{\hrule\hbox{\vrule\kern8pt\vbox{\kern8pt
\hbox{$\displaystyle #1$}\kern8pt}\kern8pt\vrule}\hrule}}

\def\IB{\relax\hbox{$\inbar\kern-.3em{\rm B}$}}
\def\IC{\relax\hbox{$\inbar\kern-.3em{\rm C}$}}
\def\ID{\relax\hbox{$\inbar\kern-.3em{\rm D}$}}
\def\IE{\relax\hbox{$\inbar\kern-.3em{\rm E}$}}
\def\IF{\relax\hbox{$\inbar\kern-.3em{\rm F}$}}
\def\IG{\relax\hbox{$\inbar\kern-.3em{\rm G}$}}
\def\IGa{\relax\hbox{${\rm I}\kern-.18em\Gamma$}}
\def\IH{\relax{\rm I\kern-.18em H}}
\def\IK{\relax{\rm I\kern-.18em K}}
\def\IL{\relax{\rm I\kern-.18em L}}
\def\IP{\relax{\rm I\kern-.18em P}}
\def\IR{\relax{\rm I\kern-.18em R}}
\def\IZ{\relax\ifmmode\mathchoice
{\hbox{\cmss Z\kern-.4em Z}}{\hbox{\cmss Z\kern-.4em Z}}
{\lower.9pt\hbox{\cmsss Z\kern-.4em Z}} {\lower1.2pt\hbox{\cmsss
Z\kern-.4em Z}}\else{\cmss Z\kern-.4em Z}\fi}

\def\II{\relax{\rm I\kern-.18em I}}

\def\CA {{\cal A}}

\begin{document}

\hfill CERN-PH-TH/2005-024

\hfill NRCPS-HE-2005-13

\vspace{5cm}
\begin{center}
{\LARGE Tensionless Strings,\\
Correspondence with SO(D,D) Sigma Model

}

\vspace{2cm}
{\sl  G. Savvidy$ $\footnote{ {\footnotesize\it E-mail: georgios.savvidis@cern.ch}}\\
\bigskip
{\sl   CERN Theory Division, CH-1211 Geneva 23, Switzerland}\\
{\sl   Demokritos National Research Center,GR-15310 Athens, Greece}
}
\end{center}
\vspace{60pt}

\centerline{{\bf Abstract}}

\vspace{12pt}

\noindent
String theory with perimeter action is tensionless  by its geometrical nature
and has pure massless spectrum of higher spin gauge particles. I demonstrate that
liner transformation of the world-sheet fields defines a  map to the
SO(D,D) sigma model equipped by additional Abelian constraint,
which breaks SO(D,D) to a diagonal SO(1,D-1). The effective tension is equal to
the square of the dimensional coupling constant  of the perimeter action.
This correspondence allows to view the perimeter action as a "square root"
of the Nambu-Goto area action. The aforementioned map
between tensionless strings and SO(D,D) sigma model allows to introduce
the vertex operators in full analogy with the standard string theory
and to confirm the form of the vertex operators introduced earlier.


\newpage

\pagestyle{plain}

It is generally expected that high energy limit, or what is equivalent
the tensionless limit $\alpha^{'} \rightarrow \infty$,
of string theory should have massless spectrum
$M^2_N  = (N-1)/\alpha^{'} \rightarrow 0$ and should
recover a genuine symmetries of the theory \cite{Gross:1987kz,Gross:1987ar,gross}.
Of course this observation  ignores the importance
of the high genus $G$ diagrams, the contribution of which
$A_G \simeq exp\{ - \alpha^{'} s/(G +1)\}$ is exponentially
large compared to the tree level diagram \cite{Gross:1987kz,Gross:1987ar,gross}.
The ratio of the corresponding
scattering amplitudes behaves as $A_{G+1} /A_G \simeq exp\{\alpha^{'} s/G^2\}$ and
makes  any perturbative statement unreliable and  requires therefore
nonperturbative treatment of the problem
\cite{Witten:1988zd,Mende:1992pm,Mende:1989wt, Mende:1994wf, Moore:1993ns}
\footnote{The different aspects
and models of tensionless theories can be found in
\cite{deVega:1987hu,Amati:1988tn,Gasperini:1991rv,
Lindstrom:1990qb,DeVega:1992tm,Lizzi:1994rn,Bakas:2004jq,Bakas:2005sd}.}.

The tensionless model with perimeter action suggested in  \cite{geo,Savvidy:dv,Savvidy:2003fx}
does not appear as a $\alpha^{'} \rightarrow \infty$ limit of the standard string
theory, as one could probably think,
but has a tensionless character by its geometrical nature \cite{geo}.
Therefore it remains mainly unclear at the moment how these two models are connected.
However the perimeter model shares many properties with
the area strings in the sense that it has
world-sheet conformal invariance, contains the corresponding Virasoro algebra,
which is extended by additional Abelian generators. This makes mathematics
used in the perimeter model very close to the standard string theory and allows
to compute its massless spectrum, critical dimension $D_c =13$ \cite{Savvidy:dv,Savvidy:2003fx}
and to construct an appropriate vertex operators \cite{alvarez,Savvidy:2004bb}.

Comparing literally the spectrum of these two models
one can see that instead of usual exponential growing of states,
in the perimeter case we have only linear growing.
In this respect the number of states in the perimeter model is much less compared with
the standard string theory  and is larger compared with the field
theory models of the Yang-Mills type.
From this point of view it is therefore much
closer to the quantum field theory rather
than to the standard string theory. At the same time its formulation and
the symmetry structure is more string-theoretical.
Perhaps there should be strong nonperturbative rearrangement of the spectrum
in the limit $\alpha^{'} \rightarrow \infty$ before the spectrum of the area and
the perimeter strings can become close to each other.

Our aim here is to give a partial answer to these questions. As we shell see the
liner transformation of the world-sheet fields defines a  map to the
SO(D,D)~ $\sigma$-model equipped by an additional Abelian constraint,
which breaks SO(D,D) to a diagonal SO(1,D-1).
The effective string tension is equal to
the square of the dimensional coupling constant $m$ of the perimeter action
$$
\frac{1}{2\pi \alpha^{'}} = \frac{m^2}{\pi}.
$$
This relation allows to view the perimeter action as a "square root"
of the Nambu-Goto area action $m = \sqrt{1/2\alpha^{'}}$.  The mass-shell
quantization condition of the $SO(13,13) ~\sigma$-model
$$
\alpha^{'} M^{2}_N = - \alpha^{'} K^2 =  (N-1),
$$
the value of the first Casimir operator $K^2$
of the Poincar\'e algebra in 26-dimensions,
is translated  through the dictionary  into the quantization condition
for the square $W = w^{2}_{D-3}$ of the Pauli-Lubanski form $w_{D-3}$
of the Poincar\'e algebra in 13-dimensions
$$
W_N = \frac{(k \cdot \pi)^2}{m^2} = (N-1)^2 ,
$$
because as we shall see (\ref{casimirmap})
$
K^2 \vert_{in~26-dim.} = 2m~ (k \cdot \pi)\vert_{in~13-dim.}
$
The aforementioned correspondence allows to introduce
the vertex operators in full analogy with the standard string theory
and to confirm the form of the vertex operators introduced earlier in
\cite{alvarez,Savvidy:2004bb}.

The n-point scattering  amplitude of {\it fixed helicity states} $W = 0 ~ (N=1)$  in
terms of 13-dimensional momenta $k_i$ and polarizations $e_i$ is\footnote{
The $W_N$ defines fixed helicity states, when $W=0~(N=1)$ and
continuous spin representations-CSR, when $W \neq 0 ~(N\neq 1)$
\cite{wigner,brink,Savvidy:dv,Savvidy:2003fx,Mund:2004sy}.}
\be
\CA (k_1 ,e_1 ;.....; k_n ,e_n) = \int d\pi_1 ...d\pi_n e^{i e_1 \pi_1 +...+ie_n \pi_n}
\int \prod^{n}_i ~d^{2}\zeta_i
<U_{k_1,\pi_1}(\zeta_1) ... U_{k_n,\pi_n}(\zeta_n)>,
\ee
where $U_{k_i,\pi_i} $ (\ref{fixedhelicityvertex})
are fixed helicity vertex operators $ (k_i \cdot \pi_i) =0,~i=1,...,n$.
This  scattering amplitude exhibits
the important gauge invariance  with respect to the gauge transformations
\cite{Savvidy:2003fx}
\be
e_i \rightarrow e_i + k_i ~\Lambda_i(k_1 , ...k_n),
\ee
where $\Lambda_i(k_1 , ...k_n)$ are gauge parameters. This invariance
is valid only for the states which are described by the fixed helicity
vertex operator $U_{k,\pi}$ (\ref{fixedhelicityvertex}), for which
$W \sim  (k \cdot \pi)^2  = 0$.

\vspace{0.3cm}

{\underline{\it The perimeter string model} was suggested in  \cite{geo} and
describes random surfaces embedded in D-dimensional
space-time  with the following action
$$
S  ~=~m~L=
{m\over\pi}\int d^{2}\zeta \sqrt{h}~\sqrt{ \left(\Delta(h)
X_{\mu}\right)^{2}},
$$
where ~$h_{\alpha\beta}$ ~is the world-sheet metric,
~$\Delta(h)= 1/\sqrt{h}~\partial_{\alpha}\sqrt{h}h^{\alpha\beta}
\partial_{\beta} $ ~is  Laplace operator and
$m$ has dimension of mass.
There is no Nambu-Goto area term in this action.
The action has
dimension of length $L$ and the dimensional coupling constant is $m$.
Multiplying and dividing the Lagrangian by the square
root $~\sqrt{ \left(\Delta(h)
X_{\mu}\right)^{2}}$
one can represent it in the $\sigma$-model form \cite{Savvidy:dv}
\beqa\label{conaction}
S= -{1\over\pi}\int d^{2}\zeta ~\eta_{\mu\nu} \sqrt{h} h^{\alpha\beta} ~
\partial_{\alpha}\Pi^{\mu}~ \partial_{\beta} X^{\nu}  ,
\eeqa
where the operator $\Pi^{\mu}$ is
$$
\Pi^{\mu} = m {\triangle (h) X^{\mu} \over
\sqrt{(\triangle (h) X^{\mu})^{2}}}.
$$
We shall consider the model {\it B}, in which
two field variables $X^{\mu}$ and $h_{\alpha\beta}$ are independent.
The classical equation is
\be\label{motioneq}
(I)~~~~~~~~~~~~~~~~~~~~~~~~~~~~~~\triangle (h)~\Pi^{\mu} =0~~~~~~~~
~~~~~~~~~~~~~~~~~~~~~
\ee
and world-sheet energy-momentum tensor
\be\label{tensor}
(II)~~~~~~~T_{\alpha\beta} = \partial_{\{\alpha }\Pi^{\mu}  ~\partial_{\beta\}} X^{\mu}
~-~~h_{\alpha\beta}~h^{cd} ~ \partial_{c}~ \Pi^{\mu}~\partial_{d} X^{\mu}~=~0.
\ee
The operator $\Pi$ is a {\it space-like vector},
\be\label{secondcon}
(III)~~~~~~~~~~~~~~~~~~~\Theta ~\equiv ~\Pi^{\mu}\Pi^{\mu}~ -~ m^2~=0.
~~~~~~~~~~~~~~~~~~~~~~~~
\ee
The energy momentum tensor is conserved
$
\nabla^{a}~T_{ab} =0
$
and is traceless $h^{ab}T_{ab} =0$, thus we have
two-dimensional world-sheet conformal field theory with the
central charge $c =2D$ \cite{Savvidy:dv}. We have equation of motion (\ref{motioneq})
together with the primary constraint equations (\ref{tensor})
and (\ref{secondcon}), the secondary
constraints have the form $\Theta^{1,0}= \Pi \partial_+\Pi,~~
\Theta^{0,1}= \Pi \partial_-\Pi,~~
\Theta^{1,1} = \partial_+\Pi \partial_-\Pi$
\cite{Savvidy:dv} .
The equivalent form of the action (\ref{conaction}) is \cite{Nichols:2002ux}
\be\label{gaga}
\acute{S} =-
{1\over\pi}\int d^{2}\zeta ~ \sqrt{h} h^{\alpha\beta} \{~
\eta_{\mu\nu}~\partial_{\alpha}\Pi^{\mu}~ \partial_{\beta} X^{\nu}~
+ ~ \omega_{\alpha\beta} ~ ( \Pi^{2} -m^2)~\},
\ee
where the $\Pi^{\mu}$ field is now an independent variable and the $\omega_{\alpha\beta}$
are a Lagrange multipliers. The system of equations which follows from
$\acute{S}$
\beqa\label{orig}
\triangle (h)~ \Pi^{\mu} =0,~~~~~~~~
\triangle (h)~ X^{\mu} - 2 h^{\alpha\beta} \omega_{\alpha\beta} \Pi^{\mu} =0,~~~~~~~~
\Pi^{\mu}\Pi_{\mu}~ = m^{2}
\eeqa
is equivalent to the original equations (\ref{motioneq}) and (\ref{secondcon}) and
the corresponding new energy momentum  tensor $\acute{T}_{\alpha\beta}$ acquires
an additional term which depends only on the field $\Pi$
\be\label{totaltensor}
\acute{T}_{\alpha\beta} = T_{\alpha\beta} ~
+ ~(\omega_{\alpha \beta}  ~-~{1\over 2}~h_{\alpha\beta}~h^{\gamma\delta} ~
\omega_{\gamma\delta})(\Pi^2 -1),
\ee
where $h^{ab}\acute{T}_{ab} =0$. The central
charge $c=2D$ of the Virasoro algebra remains untouched and demonstrates
the absence of additional contributions  to the central charge due to
the primary and  secondary constraint (\ref{secondcon})
(see also \cite{Mourad:2004fg} for alternative calculation).

\vspace{0.5cm}

{\underline{\it  Correspondence with the SO(D,D) $\sigma$-model }}.
Let us introduce the new variables as follow
\be\label{transformation}
{1\over m^2}\Pi^{\mu} = {1\over \sqrt{2}}(\Phi^{\mu}_{1} + \Phi^{\mu}_{2}),~~~~~
X^{\mu} = {1\over \sqrt{2}}(\Phi^{\mu}_{1} - \Phi^{\mu}_{2})
\ee
then the action (\ref{conaction}) will take the form
\beqa\label{dualaction}
S = -{m^2 \over 2\pi}\int d^{2}\zeta ~~\eta_{\mu\nu} \sqrt{h} h^{\alpha\beta}
(~\partial_{\alpha}\Phi^{\mu}_{1}
~ \partial_{\beta} \Phi^{\nu}_{1} - ~\partial_{\alpha}\Phi^{\mu}_{2}
~ \partial_{\beta} \Phi^{\nu}_{2}).
\eeqa
If one considers the $2D$ dimensional target space with the combined coordinates
$$
\Phi^{M} = (\Phi^{\mu_1}_{1},\Phi^{\mu_2}_{2}), ~~~~~M=1,...,2D
$$
and fully symmetric Lorenzian signature space-time metric with D pluses and D minuses
\be
\eta^{M N} =   \left( \begin{array}{l}
  \eta^{\mu_1 \nu_1}~~~~~~~~~ \\
  ~~~~~~~~~~~-\eta^{\mu_2 \nu_2}
\end{array} \right) = \left( \begin{array}{l}
  -,+,...,+ ~~~~~~~~~~ \\
  ~~~~~~~~~~~~~~~~~~+,-,...,-
\end{array} \right)
\ee
then the action (\ref{dualaction}) will have formal
interpretation in terms of  $\sigma$-model being defined on a 2D dimensional
target space with the symmetry group $SO(D,D)$
\be
S ~=~ -{m^2 \over 2\pi}\int  d^{2}\zeta ~~\eta_{MN} \sqrt{h} h^{\alpha\beta}
~\partial_{\alpha}\Phi^{M}
~ \partial_{\beta} \Phi^{N}.
\ee
From this expression of the action we can deduce that the effective string
tension $T_{eff}$ is equal to the square of the mass m
\be\label{squareroot}
\frac{1}{2\pi \alpha^{'}} = \frac{m^2}{\pi} .
\ee
The last relations allow to view the tensionless string theory, which is
defined by the perimeter action (\ref{conaction}), as a "square root
of the Nambu-Goto area action" ~ $m = \sqrt{1/2\alpha^{'}}$.
This interpretation has deep geometrical origin
because in some sense the perimeter  L, which was defined
for the two-dimensional surfaces in
(\ref{conaction}), can be consider as a square root of the surface area.
This intuitive interpretation
can be made more precise if one recalls Zenodor-Minkowski isoperimetric  inequality
\cite{minkowski,bla},
which tells that $L^2 \geq 4\pi S $, with the equality taking place only for a
sphere.

The crucial constraint  (\ref{secondcon}) will take the form
$m^2(\Phi_{1} + \Phi_{2})^2~ = 2$
and brakes $SO(D,D)$ group of fully symmetric space-time $M^{D,D}$
down to the diagonal group$SO(1,D-1)$
of the standard space-time $M^{1,D-1}$ with one time coordinate
$$
SO(D,D) \rightarrow  diag ~~SO(1,D-1), 
$$
as one can see from the component form of the above
constraint\footnote{The SO(D,D)
signature allows the D light-cone coordinates
$\Phi_{\pm} =\Phi^{0}_{1} \pm \Phi^{0}_{2},~
\vec{\Phi}_{\pm} =\vec{\Phi}_{1} \pm  \vec{\Phi}_{2} $.}
\be\label{equivalentsecondcon}
-(\Phi^{0}_{1} + \Phi^{0}_{2})^2 + (\vec{\Phi}_{1} + \vec{\Phi}_{2})^2 =
-\Phi_{+}^2 +  \vec{\Phi}_{+}^2 = 2/m^2 .
\ee
The Abelian constraint (\ref{equivalentsecondcon}) can be considered as a
"compactification" to a hyperboloid manifold $H^{D}$.
The $X_{\mu}$ and $\Pi_{\mu}$ fields (\ref{transformation}) are actually
light cone coordinates on $M^{D,D}$.

The energy momentum tensor (\ref{tensor}) will take the form
\be
 ~~~~T_{\alpha \beta} =  \partial_{\alpha }\Phi_{1}  ~\partial_{\beta} \Phi_{1}-
\partial_{\alpha }\Phi_{2}  ~\partial_{\beta} \Phi_{2}
~-~{1\over 2}~h_{\alpha\beta}~h^{\gamma\delta} ~ (\partial_{\gamma}~\Phi_{1}
\partial_{\delta}~\Phi_{1}- \partial_{\gamma}~\Phi_{2} ~\partial_{\delta}~\Phi_{2} ).
\ee
It is therefore clear that we should have $2D_c = 26$ and
recover the previous result   \cite{Savvidy:dv}
\be
D_c = 13.
\ee
{\underline{\it Operator Algebra and Vertexes }}
For the open strings the solution of this two-dimensional world-sheet CFT
is \cite{Savvidy:dv}:
\beqa\label{solution}
X^{\mu} = x^{\mu} +
{1\over m}\hat{\pi}^{\mu}\tau +
i\sum_{n \neq 0}  {1\over n }~ \beta^{\mu}_{n} e^{-in\tau}\cos n\sigma,\nonumber\\
\Pi^{\mu} =   m e^{\mu} +  \hat{k}^{\mu}\tau + i\sum_{n \neq 0}
{1\over n }~ \alpha^{\mu}_{n} e^{-in\tau}\cos n\sigma,
\eeqa
where~$\hat{k}^{\mu} = -i \partial/\partial x_{\mu}$ and
$\hat{\pi}^{\mu} = -i \partial/\partial e_{\mu}$ are momentum operators and
$\alpha_n$, ~$\beta_n$ are oscillators with the
following commutator relations
\be\label{algebra}
[x^{\mu},~ \hat{k}^{\nu}] = i\eta^{\mu\nu},~~~[e^{\mu},~\hat{\pi}^{\nu}]=
i\eta^{\mu\nu},~~~[\alpha^{\mu}_{n},~
\beta^{\nu}_{l}]=n~\eta^{\mu\nu}\delta_{n+l,0}
\ee
and $[\alpha,\alpha]= [\beta,\beta]=0$ (the indexes are not shown).
It is also convenient to introduce the zero momentum operators
$\alpha^{\mu}_{0}=\hat{k}^{\mu},~\beta^{\mu}_{0}= \hat{\pi}^{\mu}$~.
The appearance of the additional zero mode means that the wave function is
a function of the coordinate variables  $x^{\mu}$ and $e^{\mu}$:~
$$
\Psi_{Phys}= \Psi(x,e) .
$$
The coordinate variable $x^{\mu}$  belongs to a Minkowski space
$x^{\mu} \in M^{13}$ and $e^{\mu}$
belongs to a hyperboloid  $e^{\mu} \in~ H^{13}$ which is defined by the
relation  $e^2 = -e^{2}_{0}  + \vec{e}^{~2} = 1$ ~
(\ref{secondcon}),(\ref{equivalentsecondcon})
\be
M^{26}  \rightarrow M^{13}  \otimes H^{13} .
\ee
It was suggested therefore in \cite{Savvidy:dv} that $e^{\mu}$
should be interpreted as a polarization vector, because from the
constraint (\ref{secondcon}),(\ref{equivalentsecondcon})
it follows that  \cite{Savvidy:dv}
$$
k^2 =0,~~e\cdot k =0,~~e^2 = 1.
$$
It is important to get better idea about the algebra (\ref{algebra}).
The transformation (\ref{transformation}) naturally leads to the oscillators
$$
A^{\mu}_n= {1\over \sqrt{2}}(\alpha^{\mu}_n + \beta^{\mu}_n) , ~~
B^{\mu}_n = {1\over \sqrt{2}}(\alpha^{\mu}_n - \beta^{\mu}_n)
$$
and brings the algebra (\ref{algebra}) to the form
\beqa\label{standardoscillators}
[A^{\mu}_n ; A^{\nu}_m ] &=& +  \eta^{\mu\nu} ~n \delta_{n+m}\nonumber\\
~[B^{\mu}_n ; B^{\nu}_m ] &=& -  \eta^{\mu\nu} ~n \delta_{n+m}\nonumber\\
~[A^{\mu}_n ; B^{\nu}_m ]&=&0 ~.
\eeqa
This is a standard algebra of the oscillators with the following signature
\beqa
\eta^{\mu\nu} = (-,+,...,+)~~~\in~~SO(1,D-1);~~~~~~~
-\eta^{\mu\nu} =(+,-,...,-)~~~\in~~ SO(D-1,1).\nonumber
\eeqa
In terms of the above oscillators the "target space" coordinates
(\ref{transformation})~~ $\Phi^M
= (\Phi^{\mu_1}_{1},\Phi^{\mu_2}_{2})$ have the form:
\beqa\label{phisolution}
\sqrt{2}~\Phi^{\mu}_{1} = x^{\mu} + {1\over m} e^{\mu} + (~{1\over m^2} \hat{k}^{\mu} +
{1\over m}\hat{\pi}^{\mu} ~)~\tau +
{i\over m} \sum_{n \neq 0}  {1\over n }~A^{\mu}_{n}~
e^{-in\tau}\cos n\sigma~,\nonumber\\
\sqrt{2}~\Phi^{\mu}_{2} =  - x^{\mu} + {1\over m} e^{\mu} + (~{1\over m^2} \hat{k}^{\mu} -
{1\over m}\hat{\pi}^{\mu})~\tau +  {i\over m} \sum_{n \neq 0}
{1\over n }~B^{\mu}_{n}~
e^{-in\tau}\cos n\sigma~.
\eeqa
The above SO(D,D) $\sigma$-model interpretation of the tensionless
string theory allows to introduce
the vertex operators in the full analogy with the standard string theory case.
Indeed the vertex operator for the ground state has the form:
$$
V_K ~=~ :e^{iK\Phi}:
$$
where $K^M =(k^{\mu_1}_1 , k^{\mu_2}_2)$, $~\Phi^M
= (\Phi^{\mu_1}_{1},\Phi^{\mu_2}_{2})$ and has conformal dimension equal to
the square of the momentum $K^{M}$
\be\label{verter}
\Delta =  \alpha^{'}~ K^2 = \frac{\eta_{MN} K^{M}K^N}{2 m^2}~.
\ee
Therefore substituting the expressions for the field $~\Phi^M
= (\Phi^{\mu_1}_{1},\Phi^{\mu_2}_{2})$  in terms
of the original world-sheet fields $X^{\mu}$ and $\Pi^{\mu}$
(\ref{transformation}) we shall get
\be
V_K ~=~ :e^{ik_1 \Phi_1 ~-~ ik_2 \Phi_2}: ~=~ :e^{ikX +i\pi \Pi/m}:
\ee
where the momenta $k$ and $\pi$ are:
\be
k = {1\over \sqrt{2}}(k_1 + k_2),~~~~~~~~\pi ={1\over \sqrt{2}~m}(k_1 - k_2).
\ee
This is an interesting relation because it demonstrates how the
2D-dimensional  momentum
$K^{M}$ of the $SO(D,D)$ $\sigma$-model splits into two parts
which form the physical momentum
variable $k_{\mu}$ of the tensionless strings propagating in $13$-dimensional
Minkowski space-time  $x^{\mu} \in M^{1,D-1}$
and the momentum $\pi^{\mu}$ which is conjugate
to the polarization vector $e^{\mu} \in~ H^{1,D-1}$
$$
M_{SO(13,13)} \rightarrow M_{SO(1,12)} \otimes H_{SO(1,12)} .
$$
We can now translate the conformal dimension $\Delta$
of the ground state vertex operator $V_K$ into
the language of our momenta  $k$ and $\pi$
\be\label{casimirmap}
\Delta = \frac{\eta_{MN} K^{M}K^N}{2 m^2} = \frac{k^{2}_1 - k^{2}_2}{2 m^2} =
{(k + m \pi)^2\over 4 m^2} -
{(k - m \pi)^2\over 4 m^2} =   \frac{ (k \cdot \pi)}{m}.
\ee
This clearly confirms the form of the vertex operator and its conformal dimension
obtained earlier in \cite{alvarez}. Indeed the general form of the vertex
operators suggested in \cite{alvarez} is given by the formula
\be\label{highordervertex}
U^{\mu_1 \tilde{\mu}_1...,  ...\mu_j \tilde{\mu}_j}_{k,\pi}(\zeta)=
:\partial^{n_1}_{\zeta} X^{\mu_1}~
\partial^{\tilde{n}_1}_{\bar{\zeta}} X^{\tilde{\mu}_1}~...~
~ ...\partial^{n_j}_{\zeta} \Pi^{\mu_j}~
\partial^{\tilde{n}_j}_{\bar{\zeta}} \Pi^{\tilde{\mu}_j}~
e^{~i k \cdot X(\zeta) +i\pi \cdot \Pi(\zeta)~}: ~,
\ee
the conformal spin should be equal to
zero, therefore $n_1 +...+n_j = \tilde{n}_1 +...+\tilde{n}_j = N$.
Using the world-sheet energy momentum
operator \cite{Savvidy:dv} $T(\zeta)=-:\partial_{\zeta} X \cdot \partial_{\zeta}
\Pi:$ one can compute the anomalous dimension of the open strings vertex
operators   \cite{alvarez,Savvidy:2004bb}:
\be\label{conformaldimensions}
\Delta = {(k \cdot \pi)\over m} + N.
\ee
It must be equal to $1$ in order to describe
emission of physical states, therefore the corresponding poles are at
\be\label{conformalinvarianceconditions}
\frac{(k \cdot \pi)}{m}  =1 - N, ~~
\ee
which can be translated to the mass-shell condition in the
$\sigma$-model side: $\alpha^{'} K^2 = 1-N$.

Let us discuss what these relations mean.
The physical meaning of the invariant $k \cdot \pi$ is given by
$W = w^{2}_{D-3}$ - square of the Pauli-Lubanski form $w^{\mu_1 ,...\mu_{D-3} }_{D-3} \sim
\epsilon^{\mu_1 ,...\mu_{D-3},\nu\lambda\rho}
k_{\nu}~ M_{\lambda\rho}$ of the Poincar\'e algebra
on $M^{13}$, that is \cite{Savvidy:dv,Savvidy:2003fx}
\be\label{W}
W= \frac{(k \cdot \pi)^2}{m^2} .
\ee
From (\ref{conformalinvarianceconditions}),(\ref{W}) we  conclude
that on the level N the value of the
square of the Pauli-Lubanski form is equal to
\be
W_N = (1-N)^2 ,
\ee
As it is well known it defines fixed helicity states, when $W=0$ and
continuous spin representations-CSR, when $W \neq 0$
\cite{wigner,brink,Savvidy:dv,Savvidy:2003fx,Mund:2004sy},
therefore all excited states, including the ground state N=0, realize a
continuous spin representations of the massless little group $SO(11)$
and only  $N=1$ state
realizes the fixed helicity representations \cite{alvarez,Savvidy:2004bb}.
The corresponding vertex operator  $U_{k,\pi}$  ~(
$N= n = \tilde{n} =1$)~ in closed strings case is
\beqa\label{fixedhelicityvertex}
U_{k, \pi} = \xi_{MN} :\partial_{\zeta} \Phi^{M} ~\partial_{\bar{\zeta}}\Phi^N ~
~e^{~i K \Phi~}:~=~ \omega U_{XX}+2\varphi U_{X\Pi}+\chi U_{\Pi\Pi},
\eeqa
where in 26-dim. $K^{M} \xi_{MN} =0$ or being translated  through our dictionary
into 13-dim. it will take the form already found in
\cite{alvarez,Savvidy:2004bb}:~ $\pi^{\mu} \omega_{\mu\nu } +
\varphi_{\nu \mu} k^{\mu} =0,~\pi^{\mu}\varphi_{\mu\nu } + k^{\mu} \chi_{\mu\nu } =0$.~
The $U_{k,\pi}$ operators are of the essential importance, because for them
$W \sim (k \cdot \pi)^2 =0$, and they create fixed helicity
massless gauge particles \cite{pauli,singh,fronsdal,Ferrara:1998jm,Haggi-Mani:2000ru,
Sundborg:2000wp,Witten:2000,Bengtsson:1983pg,Bengtsson:1983pd,Vasiliev:1999ba,Sezgin:2001zs,
Francia:2002aa}.

In conclusion I would like to thank  Luis  Alvarez-Gaume,
Ignatios  Antoniadis, Ioannis Bakas, Lars  Brink
and Kumar Narain for stimulating discussions and
CERN Theory Division for hospitality.

\vfill
\end{document}